\documentstyle[12pt]{article}

\def\gtwid{\raise.3ex\hbox{$>$\kern-.75em\lower1ex\hbox{$\sim$}}}
\def\ltwid{\raise.3ex\hbox{$<$\kern-.75em\lower1ex\hbox{$\sim$}}}

\def\o{\omega}
\def\obar{\bar{\omega}}
\def\ubar{\bar{u}}

\def\la{\langle}
\def\ra{\rangle}
\def\pastI{{\cal I}^-}
\def\futI{{\cal I}^+}
\def\cmax{C_{\rm max}}
\def\cmin{C_{\rm min}}

\begin{document}
\title{Black Hole Radiation in the Presence of \\a Short Distance
Cutoff}
\author{Ted Jacobson\thanks{jacobson@umdhep.umd.edu}\\
Department of Physics, University of Maryland\\
College Park, Maryland 20742\\
and\\
Institute for Theoretical Physics\\
University of California, Santa Barbara, CA 93106}
\date{March, 1993}
\maketitle

\vspace{-12cm}
\begin{flushright}
UMDGR93-32\\NSF-ITP-93-26\\hep-th/9303103
\end{flushright}
\vspace{10cm}

\begin{abstract}
A derivation of the Hawking effect is given which avoids
reference to field modes above some cutoff frequency
$\omega_c\gg M^{-1}$ in the free-fall frame of the black
hole. To avoid reference to arbitrarily high frequencies, it
is necessary to impose a boundary condition on the quantum
field in a timelike region near the horizon, rather than on
a (spacelike) Cauchy surface either outside the horizon or
at early times before the horizon forms. Due to the nature
of the horizon as an infinite redshift surface, the correct
boundary condition at late times outside the horizon cannot
be deduced, within the confines of a theory that applies only
below the cutoff, from initial conditions prior to the formation of
the hole. A boundary condition is formulated which leads to the
Hawking effect in a cutoff theory.
It is argued that it is possible the
boundary condition is {\it not} satisfied, so that the
spectrum of black hole radiation may be significantly
different from that predicted by Hawking, even without the
back-reaction near the horizon becoming of order unity relative
to the curvature.
\end{abstract}

\newpage

\section{Introduction}
\label{sec:1}

The Hawking radiation from a black hole of mass $M$ is most
copious at a wavelength of order $M$.\footnote{We use
units with $G=c=\hbar=1$.} In this sense it is a long
distance effect, whose scale is set by the mass of the hole.
Thus it is odd that all derivations of the Hawking
effect refer in some manner to arbitrarily short distances.
For instance, consider Hawking's original derivation
\cite{Hawk75}: the annihilation operator for an outgoing quantum
field mode at late times is expressed, via the free field
equations, in terms of annihilation and creation operators
for ingoing modes at early times, before the matter has
collapsed to form the hole. The thermal character of the
state at late times is then deduced from the boundary
condition specifying that the initial state is the vacuum
(or vacuum plus some excitations of finite total energy.)

The fishy thing about this derivation is that the frequency
of the ingoing modes diverges as the time of the corresponding
outgoing modes goes to infinity. This is because all of the
outgoing modes, for all eternity, originate as incoming
modes that arrive at the hole {\it before} the formation of
the event horizon. An infinite number of oscillations of the
incoming modes must thus be packed into a finite time
interval, so their frequency must diverge.

Other derivations of the Hawking effect also make reference
to arbitrarily short distances. A recent derivation by
Fredenhagen and Haag \cite{FredHaag} is based on the form of the
singularity in the two-point function
$\langle\phi(x)\phi(y)\rangle$ as $x$ approaches $y$
just outside the horizon.
Similarly, arguments based on the properties of the
correlation functions on the Euclidean continuation of the
black hole metric \cite{GibbPerry} assume that the correlation
functions have the requistite analytic behavior, which
involves the form of the short distance singularities.
Finally, arguments (for conformal fields  in two dimensions)
based on conservation of the stress-energy tensor
\cite{DavFullUn,ChristFull} assume the value of the trace
anomaly, which is the result of regulating a short distance
divergence of the theory.

Since the scale of the process is set by the mass of the
hole, it would seem that it should be possible to avoid the
role of ultra high freqencies much higher than $M^{-1}$
in deriving its existence. In a previous paper \cite{Jac} this
issue was discussed in detail, and two arguments were
offered to support this point of view, one involving the
response of accelerated particle detectors and one involving
conservation of the stress-energy tensor. These arguments
were not conclusive but they did make it plausible that the
Hawking effect would occur even if there were a Planck
frequency cutoff in the frame of free-fall observers that
fall from rest far from the hole.

It now seems a mistake to focus on a Planck frequency
cutoff, since the same arguments would support the existence
of Hawking radiation as long as the high frequency  cutoff
$\omega_c$ is much larger than $M^{-1}$.
In the present paper it will be shown how Hawking's original
analysis can be modified to avoid reference to ultra high
frequencies. This will require the use of an alternate
boundary condition, which states roughly that observers
falling freely into the black hole (starting from rest
far away) see no particles at frequencies much higher than
$M^{-1}$ but less than some cutoff $\omega_c$.
That this condition implies the existence of
black hole radiation was implicit in Hawking's original
paper \cite{Hawk75}, and was later stressed by Unruh \cite{Origin}.
One contribution of the present paper is to demonstrate
in detail how the derivation can be structured so as to
entirely avoid invoking the behavior of ultra high
frequency modes. This analysis involves several sticky technicalities,
which we have attempted to address as thoroughly as possible.

This alternate boundary condition is not an {\it initial}
condition, since it is imposed for all times. Moreover, for
the reason explained above, it cannot be derived from the
early time vacuum inital condition. It is in the nature of the
horizon as an infinite redshift surface that the state of
the outgoing field modes at low frequencies descends from
presently unknown physics at very high frequency (in the
free-fall frame). Thus the validity of
the boundary condition cannot be proved within a theory that
is only valid below some high frequency cutoff.
The possibility of justifying the boundary condition on
energetic grounds will be addressed in section \ref{sec:6}.
Our conclusion will be that it is quite possible the
boundary condition is {\it not} satisfied, so that the
spectrum of black hole radiation may be significantly
different from that predicted by Hawking, even without the
back-reaction near the horizon becoming of order unity relative
to the curvature. Violations of the boundary condition leading
to a large back-reaction also seem possible, however in such
a situation the quasi-static, semiclassical framework of
our calculations is unjustified.

The rest of the paper is organized as follows. In section
\ref{sec:2}
Hawking's original derivation is reviewed. In section
\ref{sec:3} the
role of ultra high frequencies in this derivation is
discussed, and in section \ref{sec:4}
our alternate boundary condition
is formulated and discussed in detail. It is shown in
section \ref{sec:5}
that this boundary condition implies the existence
of the usual Hawking radiation. In section \ref{sec:6} the
physical basis of the boundary condition is discussed, and the
implications of a violation of the boundary condition are studied.
Section \ref{sec:7} contains some concluding remarks, and
the appendices contain technical material needed in
the rest of the paper.

\section{Hawking's reasoning}
\label{sec:2}

In this section Hawking's original derivation \cite{Hawk75}
of black hole  radiation from a non-rotating, uncharged
black hole will be reviewed. We use Wald's formulation
\cite{Wald75} in terms  of individual wavepackets, rather
than Bogoliubov transformations between orthonormal bases,
because selection of a complete basis is distracting and
unnecessary for our our purposes.

Consider an outgoing positive frequency wavepacket $P$ at late
times far from the black hole, centered on
freqency $\obar$ and retarded time $\ubar$.
(The retarded time coordinate is
defined in Appendix A.) Suppose $P$ is normalized in the
Klein-Gordon norm, so the annihilation operator for this
wavepacket is given by
\begin{equation}
a(P)=\langle P,\Phi\rangle,
\label{2.1}
\end{equation}
where the bracket
notation denotes the Klein-Gordon (KG) inner product. (See
Appendix B for the definition of the KG inner product,
and Appendix C for a discussion of this
characterization of annihilation and creation operators.)
We are interested in the state of the quantum
field ``mode" corresponding to this wavepacket.
This is partly\footnote{For simplicity we focus on the
expectation value of the number operator. In fact, the form
of the annihilation operator $a(P)$ discussed below implies
also the true thermal nature of the state. (See for example
\cite{Wald75,DeWitt,Sork}.)}
characterized by the expectation value of the number operator,
\begin{equation} \la N(P)\ra
=\la\Psi|a^{\dagger}(P)a(P)|\Psi\ra.
\end{equation}
Using the field equation $\nabla^2\Phi=0$, this number operator
can be expressed  in terms of operators whose expectation
values are fixed by initial conditions or other assumptions
on the properties of the state $|\Psi\ra$.

Propagating the wavepacket $P$ backwards in time, it
breaks up into a ``reflected piece" $R$ that scatters
off the curvature outside the matter and out to past null
infinity ${\cal I}^-$, and a ``transmitted" piece $T$
that propagates back through the collapsing matter and
then out to ${\cal I}^-$. (See Fig. 1.)
The original wavepacket $P$ can be
exressed as the sum of these two solutions, as
\begin{equation} P=R+T,
\end{equation}
and the annihilation operator for $P$ (\ref{2.1}) can thus
be decomosed as
\begin{equation} a(P)=a(R)+a(T).
\label{2.4}
\end{equation}
Since both the wavepackets and the field operator satisfy
the wave equation,
the KG inner products in (\ref{2.4}) are conserved, and can
therefore be evaluated on any Cauchy hypersurface.

Because of time translation invariance in the part of the
spacetime exterior to the matter, the reflected packet $R$
consists of the same frequencies with respect to
the Schwarzschild time
coordinate at $\pastI$ as the packet $P$ at future null
infinity $\futI$. Thus the operator
$a(R)=\langle R,\Phi\rangle$ is an annihilation operator
for an incoming wavepacket centered on frequency $\obar$.
Assuming that this mode of
the quantum field started out in its ground state, we have
\begin{equation} a(R)|\Psi\ra =0,
\end{equation}
so that
\begin{equation} \la N(P)\ra =\la\Psi|a^{\dagger}(T)a(T)|\Psi\ra.
\label{2.6}
\end{equation}

At $\pastI$ the transmitted packet $T$ is composed of both
positive and negative frequency components with respect to the
asymptotic Schwarzschild time,
\begin{equation}
T=T^{\scriptscriptstyle (+)}+T^{\scriptscriptstyle (-)},
\end{equation}
and we have the expansion
\begin{equation} a(T)=a(T^{\scriptscriptstyle (+)})-
a^{\dagger}(T^{\scriptscriptstyle (-)}{}^*).
\label{2.8}
\end{equation}
Thus $a(T)$ is a combination of annihilation and creation
operators for incoming wavepackets at $\pastI$.
Assuming that both the positive frequency part and the complex
conjugate of the negative frequency part of the packet $T$
started out in their ground states, we have
\begin{equation}
a(T^{\scriptscriptstyle (+)})|\Psi\ra=0\qquad\qquad\qquad
a(T^{\scriptscriptstyle (-)}{}^*)|\Psi\ra=0.
\label{2.9}
\end{equation}
Thus, using (\ref{2.6}), (\ref{2.8}), (\ref{2.9}) and the
commutation relation between
annihilation and creation operators (49,50) we have
\begin{equation} \la N(P)\ra =
-\la T^{\scriptscriptstyle (-)},T^{\scriptscriptstyle (-)}\ra.
\end{equation}
For a wavepacket with a spread of frequencies
$\Delta\o\ll\kappa$, the Klein-Gordon norm of the negative frequency
packet $T^{\scriptscriptstyle (-)}$ can be evaluated in terms of
the norm of $T$ as described in Appendix D, and one finds,
using (57),
\begin{equation}
-\la T^{\scriptscriptstyle (-)},T^{\scriptscriptstyle (-)}\ra
=\la T,T\ra (\exp(2\pi\obar/\kappa)-1)^{-1}.
\label{2.11}
\end{equation}
This is just what the emission would be from a body at
temperature $\kappa/2\pi=1/8\pi M$, for a mode of energy $\obar$
with absorption coefficient $\la T,T\ra$.
\section{Ultra high frequencies}
\label{sec:3}

The difficulty with this analysis is that at past null infinity,
the incoming packet $t$ consists of extremely high frequency
components, whose frequency (with respect to the asymptotic rest
frame of the hole) grows as $\sim\exp(\ubar/4M)\, \o$ as the
retarded time $\ubar$ of the outgoing wavepacket goes to
infinity. This exceeds Planck frequency $\o_P$ for $\ubar>4M
\ln(\o_P/\o)$, that is, after only several light
crossing times for the hole. That the frequency diverges in
some such manner is immediately evident from inspection of
Fig. 1. An infinite amount of time at infinity corresponds
to the interval between any finite $u$ and the horizon at
$u=\infty$. The correspondingly infinite number of field
oscillations must all be packed into the finite range of
advanced times between some $v$ and $v_0$, the advanced time of
formation of the horizon.

It is unsatisfactory from a physical point of view to base
the prediction of black hole evaporation on an assumption
that involves the behavior of arbitrarily high frequency
modes. We are ignorant of what physics might look like at
those high frequencies or corresponding short distances.
In order to be confident of the prediction of Hawking
radiation, one should formulate a derivation that avoids
this ignorance while invoking only known physics---or at
least only more reasonable extrapolations of known physics.

It is not the unknown physics of high energy
{\it interactions} that we are concerned about here.
Although we are dealing with incoming
wavepackets with arbitrarily high frequency relative to
the frame of the collapsing matter that forms the black
hole, there is no interaction between these wavepackets
and the collapsing matter. The reason is that these incoming
wavepacket modes are in their {\it ground state}, so there
is nothing for the collapsing matter to interact with.

What we are concerned about is the need to assume that
the physics is Lorentz invariant under arbitrarily large boosts.
Assuming Lorentz invariance, one can of course argue
that although the frequency of the transmitted wavepacket $t$
grows as $\exp(\ubar/4M)$ with respect to the asymptotic rest frame
of the black hole, there is always a local Lorentz
frame in which the frequency appears as low as one wishes.
The velocity of this frame relative to the black hole
approaches the speed of light as $\ubar\rightarrow\infty$,
with a boost factor $\gamma=(1-v^2)^{-1/2}=\exp(\ubar/4M)$.

We have no observations that confirm Lorentz invariance
at the level of such arbitrarily high velocity boosts
\cite{Blok,NielPic,Will}.
Probably the highest boost factors at which Lorentz invariance
might be checked anytime soon arise in cosmic ray proton
collisions. We are basically at rest with respect to the cosmic
microwave background (CMB) radiation. Assuming Lorentz
invariance, one predicts that for proton energies greater
than about $10^{20}$ eV (relative to the CMB frame),
the head-on collision of a proton with a CMB photon can
produce a pion. This process would leave its mark on the
cosmic ray proton spectrum. If this mark is eventually
observed, it will lend support to the assumption of
Lorentz invariance that went into the
calculation.\footnote{According to Sokolsky \cite{Sokol}, it
should be possible to confirm this prediction in the coming decade.}
The boost factor here relating the CMB frame to the center of mass
frame of the collision is a ``modest" $\gamma\sim 10^{12}$.

In the black hole situation, after a retarded time interval
$\Delta u\sim 4M \ln 10^{12}\simeq 10^2 M$,
the boost factor required to transform
an incoming wavepacket to low frequency would have increased
by more than $10^{12}$. Thus the above derivation of a
steady flux of Hawking radiation depends on the assumption
of Lorentz invariance arbitarily far beyond its observationally
verified domain of validity.
\section{Cutoff boundary condition}
\label{sec:4}

To avoid the need to make assumptions regarding arbitrarily
high frequency behavior we will have to give up the attempt
to derive the properties of the state of the quantum field
at late times from the initial condition that it is the
vacuum state before the hole forms. Instead, we will formulate
a different ``boundary" condition on the state that will
still imply the existence of Hawking radiation.

The alternate boundary condition is expressed in terms of
the particle states defined by free-fall observers near the
horizon that have fallen in from rest far away from the
hole. For frequencies much higher than $M^{-1}$, these particle
states are well defined by field modes with positive
frequency with respect to the proper time of the free-fall
observers. Our boundary condition will be that outgoing,
high freqency field modes are in their ground states. How
high is ``high"? Roughly, to predict Hawking radiation
to an accuracy $\eta\ll 1$, it will suffice to assume that the
outgoing modes of free-fall frequency
$\sim \eta^{-2}M^{-1}$ are in their ground state.
The statement of the boundary condition just given is
appropriate for a {\it massless, free} field. We defer to
subsection
\ref{subsec:interact} a brief discussion of the modifications
required for a treatment of massive and/or interacting fields.

To derive this alternate boundary condition from the condition
that the {\it initial} state is vacuum requires appeal to
arbitrarily high frequency modes, for the reason discussed
earlier. Thus we make no attempt here to {\it derive} this
alternate boundary condition, but rather take it as given.
The question of physical plausibility of the condition will
be taken up in section \ref{sec:6}.

\subsection{Precise formulation of the boundary condition}
\label{subsec:precise}

Actually imposing the alternate boundary condition in terms
of the proper time of the family of free-fall observers is
somewhat complicated. Instead, shall employ the affine parameter
along radial ingoing null geodesics as the relevant ``time"
variable. This turns out to amount to the same thing near the
horizon, as will now be explained.

First note that the usual radial coordinate $r$
is an affine parameter along the radial null rays (see
Appendix A). To find the rate of change of $r$ with respect to
the proper time $\tau$ along the free-fall geodesic, note
that the quantity
$p_v=g_{v\mu}dx^\mu/d\tau=(1-{2M\over r})dv/d\tau - dr/d\tau$
is conserved, since the metric is independent of
$v$ in Eddington-Finkelstein (EF) coordinates (41).
If the geodesic starts from rest at $\infty$,
one has at infinity $dv/d\tau=1$ and $dr/d\tau=0$, so $p_v=1$.
It follows then that at the horizon $r=2M$, one has
$dr/d\tau=-1$. That is, $r$ is changing at the same rate
as the proper time.

An outgoing solution $f$ to the wave equation near the
horizon is nearly independent of $v$ in EF coordinates, since
the lines of constant $r$ are nearly null there. Along the
free-fall world line near the horizon, we therefore have
$df/d\tau\cong(\partial f/\partial r) dr/d\tau
\cong-(\partial f/\partial r)$.
Thus, for outgoing modes near the horizon,
the frequency with respect to $r$ on a constant $v$ surface
is effectively the negative of the frequency with respect to
the free-fall observers.

The particle states of our boundary condition will
correspond to wavepackets $f$ composed of field modes
on a constant $v$ null hypersurface $\Sigma$ of the form
\begin{equation}
f_{\o lm}(r,\theta,\phi)=r^{-1}\exp(i\o
r)Y_{lm}(\theta,\phi)\; .
\label{4.1}
\end{equation}
In an effort to avoid confusion I will call these {\it positive}
$r$-frequency modes, because they have positive frequency with
respect to the proper time of the free-fall
observers. We can regard the operator $a(f)=\la f,\Phi\ra$
as (proportional to) an annihilation operator for a one
particle state provided that the Klein-Gordon (KG) norm of $f$ is
positive. (This is discussed in Appendix C.)

To evaluate the Klein-Gordon inner product (44) on $\Sigma$,
we use the metric components in EF
coordinates (41) and the surface element (46) to find
$\sqrt{-g}g^{\mu\nu}d\Sigma_\nu
=-\delta_r{}^\mu\, r^2 sin\theta dr d\theta d\phi$.
Thus the KG inner product takes the form
\begin{equation}
\la f,g\ra = -{i\over2}\int d\Omega \int_0^\infty dr
\; r^2 \; (f^*\partial_r g-g\partial_rf^*).
\label{KGnorm}
\end{equation}
This shows that the modes $f_{\o lm}$ (\ref{4.1}) indeed have
positive norm for $\o>0$, as do localized wavepackets
constructed by superposing them.\footnote{It is tempting
to try to define a
full Hilbert space of one-particle states on a constant
$v$-surface using the positive $r$-frequency modes.
However, the fact that the $r$-integral runs
only over the interval $[0,\infty)$ leads to a problem with
this definition. Positive frequency modes of the form $f_{\o lm}$
and $f_{\o' lm}$ (\ref{4.1}) are {\it not} orthogonal for $\o\ne\o'$,
and linear combinations of positive frequency modes can have
negative norm. This is not a problem if one restricts
attention to wavepackets that have negligible
support near $r=0$, since for them it makes no difference whether
the $r$-integration is over $[0,\infty)$ or $(-\infty,\infty)$.
(One cannot take wavepackets of compact support since that would
be inconsistent with their being composed of purely positive
frequencies.) In any case, we will refer to only one wavepacket
at a time, with no need to consider the full Hilbert space of
one particle states.}

Our alternate boundary condition can thus be implemented as
follows. We choose to calculate the expectation value of
the number operator corresponding to wavepackets $P$
with the property that on {\it some} constant $v$ surface,
$v=v_c$, their transmitted piece $T$ has only
components with $r$-frequency $\o^{(r)}$ much higher than $M^{-1}$
but less than some cutoff frequency $\omega_c$,
\begin{equation}
\o_c>\o^{(r)}\gg M^{-1}.
\label{4.3}
\end{equation}
(If the frequency at infinity $\o$ is much greater than
$M^{-1}$, we also require $\o^{(r)}\gg\o$.)
Then, instead of propagating the transmitted
piece $T$ of the wavepacket $P$ all the way back
through the collapsing matter and out to past null infinity,
we stop when it reaches $v=v_c$.
There we decompose it into its positive and negative
$r$-frequency parts and impose the boundary condition that the
positive $r$-frequency part (and the complex conjugate of
the negative frequency part) are in their ground
states.\footnote{Although the wavepacket $P$ is completely
{\it outside} the horizon, its positive and negative
$r$-frequency parts have support both inside and outside the
horizon. (See equations (\ref{eq:tneg}), (53), (54).)}
To carry out this program, it must first be established that
there {\it exist} positive $u$-frequency wavepackets with the
property than on some surface $v=v_c$, their $r$-frequency
components satisfy (\ref{4.3}). This will be accomplished in
subsection \ref{subsec:existence} below.

\subsection{Self-consistency of the boundary condition}

Note that for a wavepacket centered
on frequency $\obar$ and retarded time $\ubar$, the surface
$v=v_c$ must necessarily move to the future as $\ubar$ grows
with $\obar$ fixed, in order to avoid the occurence of
$r$-frequency components above the cutoff frequency. Thus
our boundary condition is not being imposed on a single Cauchy
surface, so is not an ``initial" condition. This raises the question
whether our boundary condition is consistent with the field dynamics.

For simplicity, let us think of the boundary condition as being
imposed on a surface of fixed radius, $r=r_{\rm b.c.}$, just outside
the horizon.\footnote{Actually, the boundary condition
refers to the region {\it inside} the horizon as well, since
the positive and negative frequency parts have support inside
the horizon. It is therefore more accurate to think of the
boundary condition as being imposed on a {\it pair} of
surfaces of constant $r$, one just outside the horizon and
one just inside. Since the one inside is {\it spacelike}, no
question of consistency arises for that part of the boundary
condition.}
This surface is timelike, so the site of
the part of the boundary condition imposed at advanced time $v$
includes, within its past, sites of parts of the condition imposed
at earlier advanced times. Is the condition imposed at $v$ consistent
with
the earlier ones?

The boundary condition refers to the
state of outgoing modes with $r$-frequency $\o^{(r)}$ in the range
$\o_c>\o^{(r)}\gg M^{-1}$. The modes of frequency $\o_c$ come from two
sources: modes that propagate out from yet closer to the horizon
with yet higher frequencies, and modes that have scattered off
the geometry.
The state of the former modes can be freely specified, since they are
above the cutoff until they reach advanced time $v$ and hence no
condition at all is imposed on them until then.
Thus there is enough freedom to consistently assign the
state of the outgoing modes at $\o_c$. But one may still ask
if the gound state boundary condition is the {\it
appropriate} one, in view of the contributions from the
modes that have backscattered. For instance, some Hawking
radiation can scatter back towards the
hole and then scatter again out from the hole, apparently leading
to some non-zero occuption number in an outgoing mode that the
boundary condition assigns to its ground state. The scattering
amplitude for these modes in this region of the spacetime is very
small however, so such processes should affect the
state only very little.

Now let us consider the modes with frequency {\it less} than
the cutoff. The state of these modes can not really be
independently specified, since they can be traced back
(primarily) to modes yet closer to the horizon with
frequency $\o_c$, on which a (ground state) boundary
condition has already been imposed. Thus the state of the
modes with frequency $\o^{(r)}<\o_c$ must be {\it calculated},
not assigned. In fact, it follows from the argument in
section \ref{sec:5} that no modes are excited while they are
propagating close to the horizon; it is not until they climb
away significantly (on the scale of $M$) that the presence
of Hawking radiation becomes apparent in the free-fall
frame.

Thus it appears not inconsistent to impose our
ground state boundary condition, at least to the order of precision
of our calculations. Note that we can really only check
self-consistency of the calculation: As shown in the next
two subsections, the unavoidable spread of the wavepackets
makes it necessary to imose a boundary condition on a wide
range of frequencies from the beginning. Then all we can do
is verify that this boundary condition is self-consistent.

\subsection{Existence of the required wavepackets}
\label{subsec:existence}

Let $p_{\o lm}$ denote the solution to the massless scalar
wave equation in Schwarzschild spacetime
that is purely outgoing at future null infinity
(and is therefore outgoing at the horizon as well), and is of
the form
\begin{equation}
p_{\o lm}=(2\pi\omega)^{-1/2}
\exp(-i\o t) r^{-1}f_{\omega l}(r)
Y_{lm}(\theta,\phi),
\end{equation}
with
\begin{equation}
f_{\o l}(r)=\left\{
\begin{array}{l}
e^{i\o r^*}+A_{\o l}e^{-i\o r^*}\qquad\qquad
{\rm as}\; \;  r^*\rightarrow +\infty\\
B_{\o l}e^{i\o r^*}\qquad\qquad {\rm as}\; \;
r^*\rightarrow -\infty\; ,
\end{array}\right.
\end{equation}
where $r^*$ is the tortoise coordinate defined in eqn.
(\ref{tortoise}). These modes are normalized according to
$\la p_{\o lm},p_{\o' l'm'}\ra =
\delta(\o-\o')\delta_{ll'}\delta_{mm'}$.
Using these modes, we seek to construct wavepackets that
satisfy the condition (\ref{4.3}) restricting the $r$-frequency
components on a constant $v$ surface, $v=v_c$.

The wavepackets we will employ are of the following form:
\begin{equation}
P_{\obar\ubar lm}=
{\cal N}
\int_{\obar}^{\obar+\Delta\o} d\o
\, B_{\o l}^{-1}
\exp(i\o\ubar)\, p_{\o lm}.
\label{5.1}
\end{equation}
$P_{\obar\ubar lm}$ is a unit norm, positive
$t$-frequency wavepacket centered on frequency $\obar+{\Delta\o\over2}$.
${\cal N}$ is a normalization factor, and
the factor $B_{\o l}^{-1}$ (inverse of the transmission
amplitude) is included in the integrand so that we
will have control over the spread of the part of the packet
near the horizon. The wavepacket $P_{\obar\ubar lm}$ is defined by
its (purely outgoing) behavior at $\futI$
and the fact that it vanishes on
the horizon. Alternatively, propagating it backwards in time from
$\futI$ as in section \ref{sec:2},
one sees that it is generated by data on a Cauchy hypersurface formed
by a constant $v$ surface $v=v_c$ together with the part of
$\pastI$ that lies to the future of $v_c$. The wavepacket
generated by the data at $v=v_c$ alone will be called the
``transmitted packet" $T_{\obar\ubar lm}$, and that
generated by the data at $\pastI$ will be called the
``reflected packet" $R_{\obar\ubar lm}$. Thus we have
$P_{\obar\ubar lm}=
T_{\obar\ubar lm}+R_{\obar\ubar lm}$.

For each $\obar$ and for $\ubar$ sufficiently long after the
collapse that formed the black hole, one can always choose
$v_c$ sufficiently far in the past so that
$T_{\obar\ubar lm}$ is concentrated near the horizon. In
this case, the asymptotic form
$f_{\o l}\cong B_{\o l}\exp(i\o r^*)$
can be accurately substituted in the integrand
(\ref{5.1}) and one obtains
\begin{equation}
T_{\obar\ubar lm}={\cal N}
(2\pi)^{-1/2} r^{-1}Y_{lm}(\theta,\phi)
\int_{\obar}^{\obar+\Delta\o} d\o \; \o^{-1/2}\;
\exp(i\o(\ubar-u)).
\label{5.11}
\end{equation}

This transmitted wavepacket is localized in retarded time
$u$, centered roughly on $\ubar$, with a spread
$\Delta u\simeq 8\pi/\Delta\o$. More precisely,
the spread of $T_{\obar\ubar lm}$ in $u$ is
of course infinite, but the packet is well localized in the
following sense.\footnote{The wavepacket
$P_{\obar\ubar lm}$ at $\futI$
does not have the same width in $u$ as does
$T_{\obar\ubar lm}$ at $v_c$. The wavepacket is somewhat
dispersed, since the different frequency
components have unequal transmission amplitudes.
We included the factor $B_{\o l}^{-1}$ in the
definition (\ref{5.1}) of $P_{\obar\ubar lm}$ so that
our packet would be well localized at $v_c$; it will not
bother us that $P_{\obar\ubar lm}$ is not as well localized
at $\futI$.}
After carrying out the
angular integrals the KG norm (\ref{KGnorm}) of the packet
(\ref{5.11})
calculated at $v=v_c$ reduces to a numerical factor times an
integral over $x$ of the quantity $(\sin x/x)^2$, where
$x=\Delta\o(u-\ubar)/2$.
One can show that
$\int_0^y(\sin x/x)^2\, dx=(\pi/2)[1-(1/\pi y)+O(y^{-2})]$.
Thus, defining $\eta$ as the fraction of the full norm omitted in a
range $\Delta u$, one
has $\eta\simeq 1/\pi y = 4/\pi \Delta\omega\, \Delta u$, or
\begin{equation}
\eta\simeq 1/\Delta\omega\, \Delta u\, .
\label{eq:eta}
\end{equation}

In Hawking's paper \cite{Hawk75}, wavepackets of the
form (\ref{5.1}) (without the factor of $B_{\o l}^{-1}$)
were also employed, however
$\Delta\o$ was chosen very small compared with
the surface gravity $\kappa=1/4M$, so that the wavepackets
relevant to the black hole radiation would be very peaked
in frequency, thus simplifying the analysis. From our point
of view, the difficulty with this is that
such a packet cannot be squeezed close enough to the horizon
without containing $r$-frequencies above the cutoff $\o_c$.
In fact, one must take $\Delta\o\gtwid \kappa$, and to maximize
the precision of our derivation one should take
$\Delta\o\sim \sqrt{\o_c\kappa}$, as will now be shown.

\subsection{Precision of the derivation}
\label{subsec:precision}

The precision of the derivation we will give is limited by the
fact that the wavepackets will not be infinitely squeezed up against
the horizon. The resulting ``error" is of order
$C_{\rm max}\equiv (1-2M/r_{\rm max})$, where $r_{\rm max}$ is the
largest value of $r$ to occur in the wavepacket.\footnote{Actually, since
only a fraction of the wavepacket is located at
$r\sim r_{\rm max}$, with
the rest at smaller values of $r$, the error is somewhat smaller.
To keep the crude analysis that follows from getting too complicated,
we will simply make the conservative error estimate using the largest
value of $r$.}
Of course, strictly speaking, $r_{\rm max}=\infty$, but a fraction
$(1-\eta)$ if the wavepacket is contained within a smaller range
of $r$ values, given by $\Delta u\simeq 1/\eta\Delta\omega$.
Thus to minimize the errors we should minimize the combined error due to
the fraction $\eta$ of the wavepacket beyond $r_{\rm max}$, and
due to $C_{\rm max}$ not vanishing.
To carry out this minimization calculation, we must express
$\cmax$ as a function of $\eta$ and $\Delta\o$,
and minimize the error function
\begin{equation}
E^2(\eta,\Delta\o)\equiv\eta^2+\cmax^2(\eta,\Delta\o)\, .
\label{eq:E}
\end{equation}

The relation between $u$ and $r$ at constant $v$ is given
(cf. (\ref{tortoise}),(\ref{uv})) by
$\partial u/\partial r|_v=-2(1-{2M\over r})^{-1}= -2C^{-1}$,
where $C=(1-{2M\over r})$. It is this factor
that converts between $u$-frequency and $r$-frequency
at fixed $v$, $\o^{(r)}=-2C^{-1}\o$.
We assume that on the constant $v$ surface, the wavepacket is squeezed
very near to the horizon, since that is in any case required in order
to deduce the existence of Hawking radiation from our boundary condition.
Then we have (with $\kappa=1/4M$)
\begin{equation}\cmax/\cmin\simeq
\exp(\kappa\Delta u)
\sim \exp(\kappa/\eta\Delta\o)\, .
\label{eq:ratio}
\end{equation}
Now assuming the highest $r$-frequency present in the wavepacket is
the cutoff frequency, we have
$\o_c=\o^{(r)}_{\rm max}=\cmin^{-1}\o_{\rm max}$,
so that $\cmin=\o_{\rm max}/\o_c$.
Together with (\ref{eq:ratio}) this yields
\begin{equation}
\cmax\sim \exp(\kappa/\eta\Delta\o)\,
(\obar+\Delta\o)/\o_c\, ,
\label{eq:cmax}
\end{equation}
where we have returned to the notation
$\obar\equiv\o_{\rm min}$.
For the purposes of minimizing the error, we will consider
$\obar$ as fixed, since this is really determined by which
frequencies we want to learn about.

Already (\ref{eq:cmax}) shows us that it is not acceptable
to choose $\delta\o\ll\kappa$ as Hawking did. For instance,
suppose that $\obar\sim O(\kappa)$, so the frequencies most
copious in the Hawking radiation will be included,
and suppose that $\Delta\o=0.01\kappa$ and $\eta=0.01$.
Then we have $\cmax=\exp(10,000)\; \kappa/\o_c$,
which will be smaller than unity only if $\kappa/\o_c$ is
much smaller than we want to assume!

To minimize the error (\ref{eq:E}), we use (\ref{eq:cmax}) and
set $\partial E/\partial \eta=0$ and
$\partial E/\partial \Delta\o=0$.
Up to factors of O(1), this yields at the minimum:
\begin{equation}
\eta\sim(\Delta\o)^3/\kappa\o_c^2
\qquad {\rm and} \qquad
\cmax\sim(\Delta\o)^5/\kappa^2\o_c^3\; ,
\end{equation}
where $\Delta\o$ satisfies
\begin{equation}
\obar+\Delta\o\simeq(\Delta\o)^5/\kappa^2\o_c^2\, .
\end{equation}
As long as $\obar\ll \sqrt{\kappa\o_c}$, the solution is
given by
\begin{equation}
\Delta\o\sim \sqrt{\kappa\o_c},
\qquad \eta\sim\sqrt{\kappa/\o_c},
\qquad \cmax\sim\sqrt{\kappa/\o_c}.
\label{eq:error}
\end{equation}
Note that for such a ``minimum error" wavepacket with
$\o^{(r)}_{\rm max}=\o_c$, we have
$\o^{(r)}_{\rm min}=\cmax^{-1}\obar
\sim (\obar^2\o_c/\kappa^3)^{1/2}\; \kappa\, ,$
which will satisfy the condition $\o^{(r)}\gg \kappa$
as long as $\obar\gg(\kappa/\o_c)^{1/2}\, \kappa$.

We conclude that one can work with wavepackets with $r$-frequencies
in the required range, with a built-in imprecision of the
calculation\footnote{It may be that the derivation can be improved,
reducing the
imprecision. The wavepacket analysis employed here seems a rather
clumsy approach to the problem. The problem can also be
formulated using the approach of Fredenhagen and
Haag\cite{FredHaag}, which focuses on the behavior of the two-point
function. That approach may turn out to be more suitable for
maximizing the precision of the derivation.}
limited to an error of order $\sqrt{\kappa/\o_c}$.

\subsection{Horizon fluctuations}
\label{subsec:horfluct}

Another point that should be checked is how close to the horizon is
our boundary condition being imposed? If this is within the
expected range of quantum fluctuations of the horizon itself,
then we will not have succeeded in formulating a derivation
free of short distance uncertainties. To estimate the radius
$r_{\rm b.c.}$ at which the boundary condition is being imposed,
note that for a mode of frequency $M^{-1}$ coming from a hole of
mass $M$, we have $\o^{(r)}\sim\o_c$ when
$(1-{2M\over r})^{-1}M^{-1}\sim\o_c$, or
$r_{\rm b.c.}\sim 2M+ \l_c$, where $l_c=\o_c^{-1}$.
The scale of quantum fluctuations of
the horizon $\delta r$ can be estimated by using the Beckenstein-Hawking
entropy $S={1\over4} A/l_P^2$ and setting
$\delta S\sim 1$, which is characteristic of thermal
fluctuations about equilibrium.\footnote{This gives the same
scale as the one obtained by York using the
uncertainty priciple and the spectrum of quasinormal modes
\cite{York1}, or using the Euclidean partition
function approach \cite{York2}.}
Assuming the horizon should be treated as $N\equiv A/l_P^2$ independent
fluctuating area elements, each of area $a$ and radius $r$, we
have $\delta A\sim\sqrt{N}\delta a\sim l_P\delta r$,
so $\delta A\sim l_P^2$ gives $\delta r\sim l_P$.
Thus for a Planck scale cutoff, we are perhaps not justified
in ignoring the quantum fluctuations of the horizon in our derivation.
The simple way out is to take the cutoff length much longer than the
Planck length. This is fine until we come to discussing the
physical justification for the boundary condition, or violations
of it. It should be kept in mind that if the modes are followed all
the way back to where they are squeezed up within one Planck length of the
horizon, several grains of salt should be added to the whole analysis.

\subsection{Massive or interacting fields}
\label{subsec:interact}

In order to apply the arguments of section \ref{sec:5},
it is necessary that the propagation be governed by the
massless wave equation for a sufficiently long interval of
advanced time $v$. Thus
for a free field of mass $m$ one must impose the boundary
condition on wavepackets satisfying $\o^{(r)}\gg m$, in
addition to the condition $\o^{(r)}\gg M^{-1}$ already discussed
in section \ref{subsec:precise}. Then one finds that particles
corresponding to these wavepackets are created near the
black hole just as are massless ones, and they then propagate
away from the hole as massive particles. As long as the mass
is much less than the cutoff frequency, $m\ll\o_c$, there is
no obstruction to extending our argument to cover the case
of massive particles.

It is generally believed that the Hawking effect occurs for
interacting fields as well as for free fields, although this
has never been demonstrated explicitly. For the purposes of
determining what would be emitted by a real black hole, some
researchers \cite{Halzen} have assumed that the process  can
be divided into two stages, much as for the massive free
field just discussed. In the first stage, which takes place
very near the horizon, the dynamics of the field is governed
by the asymptotically free regime. In QCD for
example, free quarks and gluons are assumed to be radiated
with a thermal spectrum. In the second stage, as the
particles climb away from the horizon, the self-interactions
of the field become important, and the free particle states
hadronize into jets.

A direct demonstration of the validity
of this picture has never been given, although there are
various arguments that support it. Gibbons and
Perry\cite{GibbPerry} argued that the periodicity of the
Euclidean section of Schwarzschild
spacetime implies the thermal character of Hawking radiation for
interacting fields. This argument
applies only to the thermal equilibrium state on the eternal
black hole spacetime. Moreover, it rests heavily on the
assumption that a state that is regular on the horizon must
arise by analytic continuation from a state that is regular
on the (periodically identified) Euclidean section. While
this condition seems natural in some sense, it has not been
demonstrated to be necessary.

Another argument advanced in favor of thermality is that of Unruh
and Weiss\cite{UnruhWeiss}, who demonstrated that the {\it Minkowski}
vacuum of an interacting field theory is a thermal state when
viewed by a uniformly accelerating family of observers. More
precisely, correlation functions in the Rindler wedge are given
by the thermal density matrix relative to the Hamiltonian that
generates translations along the boost Killing field.
This is a purely kinematical result. It is, in a sense, a
local version of the Euclidean section argument that avoids
the need for assumptions about regularity of the
analytically continued correlation functions on the
Euclidean section. To turn it into a derivation of Hawking
radiation for interacting fields, one presumably must assume
the field is in a state that ``looks like" the Minkowski vacuum
very near the horizon, use the Unruh-Weiss result to describe
it from the point of view of the static observers as a thermal
state, and then propagate this thermal state out away from
the hole. The result will depend on the interactions and on
what state is incoming from infinity, since this would interact
with the outgoing Hawking radiation.

For weakly coupled fields one can study this process using
perturbation theory. Massless $\lambda\phi^4$ theory in a
2-dimensional black hole spacetime was studied by Leahy and
Unruh\cite{LeahUn}, who showed that for an ingoing thermal
state at the Hawking temperature, the interaction preserves
the thermal nature of the outgoing state. For an ingoing
vacuum state however, the outgoing state is {\it not}
thermal.

It does not appear to be entirely straightforward to extend
the arguments of our paper to the case of interacting
fields, since we use the linearity of the field equation
to express the annihilation operator corresponding to a
wavepacket at one time in terms of annihilation and creation
operators associated with wavepacket at another time.
In order to extend our argument, one can presumably use the
fact that for the first part of the process, as the
excitations are created, only the propagation of the field
``near the light cone" is relevant. That is, one can
presumably show that only small spacetime intervals are
involved, and thus use the fact that the correlation
functions behave like free field ones in this region, due to
asymptotic freedom. This picture of the process was
outlined by Fredenhagen and Haag in the discussion section
of \cite{FredHaag}, but to my knowledge it has never been worked
out in any detail.

\section{Hawking radiation in the presence of a cutoff}
\label{sec:5}

Having formulated in the previous section a boundary condition on
the quantum state near the horizon that refers only to modes below the
cutoff, it is now our task to determine the properties of the state
far from the hole.

\subsection{Evaluating the occupation numbers}

Suppose now that
$P_{\obar\ubar lm}= R_{\obar\ubar lm}+ T_{\obar\ubar lm}$ is
an outgoing wavepacket of
the form (\ref{5.1}), and
propagate $T_{\obar\ubar lm}$ back to a constant $v$
surface $v=v_c$ on which its $r$-frequency
components satisfy $\o_c>\o^{(r)}\gg M^{-1}$.(More
precisely, it will be composed of both positive and negative
$r$-frequency modes with frequencies in this range.)

Now we would like to evaluate the expectation value of the
number operator $N(P_{\obar\ubar lm})$, subject to our
``boundary conditions" on the quantum state. These are
that
\begin{enumerate}
\item the reflected piece $R_{\obar\ubar lm}$ is
in its ground state at $\pastI$, and
\item the positive $r$-frequency part and the
complex conjugate of the negative $r$-frequency part of the
transmitted piece $T_{\obar\ubar lm}$ are in their ground
states on the surface $v=v_c$ on which the $r$-frequency
components satisfy $\o_c>\o^{(r)}\gg M^{-1}$.
\end{enumerate}
Subject to these boundary conditions, the evaluation of
$\la N\ra$ goes through as in section \ref{sec:2} and we find
\begin{equation}
\la N(P_{\obar\ubar lm})\ra=
-\la T^{\scriptscriptstyle(-,r)}_{\obar\ubar lm},
T^{\scriptscriptstyle(-,r)}_{\obar\ubar lm}\ra
\label{6.1}
\end{equation}
where $T^{\scriptscriptstyle(-,r)}_{\obar\ubar lm}$
denotes the negative $r$-frequency part of the wavepacket
$T_{\obar\ubar lm}$, evaluated on the surface $v=v_c$.

Now the KG norm in (\ref{6.1}) cannot have the form of
(\ref{2.11}) because, as explained in sections
\ref{subsec:existence} and \ref{subsec:precision},
the spread of frequencies $\Delta\o$ in the packet
must be taken to be at least of order $\kappa$
(or even much larger in order to maximize the precision).
In order to exploit the simple formula (54)
that is applicable to
the negative $r$-frequency part of a wavepacket of the
form (\ref{5.1}) with $\Delta\o\ll\kappa$, we break up the packet
$P_{\obar\ubar lm}$ into a large number of pieces, defining
\begin{equation} P_{\obar\ubar lm}=\sum_{j=0}^{N-1} p_j
\label{6.2}
\end{equation}
\begin{equation} p_j={\cal N}
\int_{\obar+j\Delta\o/N}^{\obar+(j+1)\Delta\o/N}
d\o\; B_{\o l}^{-1}\;
\exp(i\o\ubar)\; p_{\o lm}
\label{packet}
\end{equation}
and the corresponding transmitted packets $t_j$.
The $\{p_j\}$ (and the $\{t_j\}$) are an orthogonal (but
non-normalized) set of
wavepackets, of the type used in Hawking's original
derivation when $N$ is chosen large enough so that
$\Delta\o/N\ll\kappa$.
(Note that for such large $N$, $B_{\o l}^{-1}$ does not
vary much over the range of integration in
(\ref{packet}) and can be pulled out of the integral.)

Each packet $t_j$ has a width of order $\Delta u\sim N/\eta\kappa$,
and therefore contains $r$-frequency components in the ratio
$\o^{(r)}_{\rm max}/\o^{(r)}_{\rm min}\sim e^{N/\eta}$ (see equation
(\ref{eq:ratio})). Nevertheless, the full wavepacket
$T_{\obar\ubar lm}$
contains only $r$-frequencies in the range (\ref{4.3});
the other $r$-frequency components in the $t_j$'s must
cancel in the sum
(\ref{6.2}), since the sum gives a much more localized wavepacket
(which suffers much less differential redshift).
It is important to stress that although we work with the packets
$t_j$ as a technique to evaluate the r.h.s. of (\ref{6.1}), we do
not attribute any direct physical significance or quantum
state to them.

Since extracting the negative frequency part is a linear
operation, we have
\begin{equation} \la T^{\scriptscriptstyle(-,r)}_{\obar\ubar lm},
T^{\scriptscriptstyle(-,r)}_{\obar\ubar lm}\ra=
\sum_{j,k}\la t^{\scriptscriptstyle (-,r)}_j,
t^{\scriptscriptstyle (-,r)}_k\ra.
\label{6.4}
\end{equation}
To evaluate the KG inner products
$\la t^{\scriptscriptstyle (-,r)}_j,
t^{\scriptscriptstyle (-,r)}_k\ra$
we would like to make use of the expression (54) for
$t^{\scriptscriptstyle (-,r)}_j$ as a linear combination of
$t_j$ and the ``time reflected" packet $\widetilde {t}_j$.
That is, we would like to use the formula
\begin{equation}
t^{\scriptscriptstyle(-,r)}_j=c_{-}
(e^{-\pi\o_j/\kappa}t_j+\widetilde{t}_j),
\label{eq:tneg}
\end{equation}
where
\begin{equation}
c_{-}=
e^{-\pi\o_j/\kappa}(e^{-2\pi\o_j/\kappa}-1)^{-1}
\end{equation}
and
\begin{equation}
\o_j=\obar+j\Delta\o/N.
\end{equation}
Now this expression for $t^{\scriptscriptstyle(-,r)}_j$
was derived in Appendix D assuming that the wavepacket
$t_j$ is squeezed close to the horizon. However, although
$T_{\obar\ubar lm}$ is squeezed close to the horizon, the
individual wavepackets $t_j$ may not be, since their width
$\Delta u$ is much larger than that of $T_{\obar\ubar lm}$.

Fortunately this is not a problem, for the following reason.
Since the KG norm is conserved, we can choose to evaluate
(\ref{6.1}) on an earlier surface $v<v_c$, on which not only
$T_{\obar\ubar lm}$ but also all the $t_j$ are squeezed
close to the horizon. Moreover, the negative $r$-frequency
part of $T_{\obar\ubar lm}$ at $v=v_c$ evolves to the
negative $r$-frequency part at $v<v_c$. This is because
$T_{\obar\ubar lm}$ is a function of $r$ only through $u$
in this region. Since $u=v-2r-4M\ln({r\over 2M}-1)$,
a shift in $v$ is equivalent to a scaling of $r$ near the
horizon (where the logarithm is dominant)
by a linear transformation $r\rightarrow ar+b$,
which leaves the negative
$r$-frequency part unchanged. This means we can evaluate the
r.h.s. of (\ref{6.1}) at a surface upon which the $t_j$ {\it are}
sufficiently sqeezed to justify use of the formula
(\ref{eq:tneg}).

The cross-terms in the sum (\ref{6.4}) vanish, since
$\la t_j,t_k\ra=\la\widetilde{t}_j,\widetilde{t}_k\ra=
\la t_j,\widetilde{t}_k\ra=0$ for $j\ne k$.
The diagonal terms are given by the result (57), so we have
finally
\begin{equation} \la N(P_{\obar\ubar lm})\ra=
\sum_j \la t_j,t_j\ra(\exp(2\pi\o_j/\kappa)-1)^{-1}.
\end{equation}
This is just what the expected occupation number would be
for a wavepacket mode of the form (\ref{5.1}) (equivalently
(\ref{6.2})) emitted from a body at temperature $\kappa/2\pi$
with absorption coefficients
$\la t_j,t_j\ra$
for the component wavepackets $p_j$.

\section{Physics of the boundary condition}
\label{sec:6}

In this section we take up the question of
whether there is any way to argue that the
boundary condition is in fact satisfied.
Recall that because of the gravitational redshift
there is no way, within a cutoff theory, to
{\it derive} the quantum state of the high
frequency outgoing modes just outside
the horizon.
The natural expectation would be that they will be in their
``free-fall" ground state, because from their
point of view, there is nothing special about the
horizon and they are merely propagating along just
as they would in flat spacetime. The problem with
this line of reasoning is that it ignores the very
question we are trying to address: does the fact
that these modes have been redshifted down from
physics above any cutoff scale leave an imprint
on their quantum state?

\subsection{Is this a one-scale problem?}

Together with the presence of the horizon,
the absence of any scale other than the size of
the black hole is really the essence of the
Hawking effect.
One can almost deduce the Hawking result
from the fact that the boundary condition
introduces no length scale other than the
Schwarzschild radius into the problem.
In our form, the
boundary condition states that
field modes near the horizon with $r$-frequencies satisfying
$\o_c>\o^{(r)}\gg M^{-1}$ are in their ground state. (Since we impose
this boundary condition for all times, no condition need be
imposed on modes with $\o^{(r)}>\o_c$.) This ground state is a pure
state, however the state of every mode outside the horizon
is correlated to that of another mode inside the horizon.
When only the field outside is
accessible, there is missing correlation information.
An observer far from the hole can never determine
the state of the modes inside the horizon, so the
relative phases of the states of all those outgoing modes at
infinity that emerged from the region of the
horizon are completely unknown. The state thus cannot be a
pure state, but is rather one in which
the missing information must be {\it maximized} in
some sense. A maximum entropy state is a thermal
one, so the state of the outgoing modes should appear
thermal (modulo absorption coefficients) far from the hole. Since
the cutoff $\o_c$ plays no quantitative role
in the problem as formulated, the only scale is
$M$, so the temperature must be proportional to
$1/M$. Calculation shows it to be $T_H=1/8\pi M$.

In the formulation where the boundary condition
is imposed in the asymptotic past, the insensitivity of
the black hole radiation to the details of the inital state
before the hole forms follows from the nature of the horizon as
an infinite redshift surface: the more time
passes, the higher the frequency of the relevant
ingoing modes. In the limit of infinite time,
all that matters is the fact that the infinitely
high frequency modes are assumed to be initially
in their ground state.

But what if one does not {\it assume} that
physics is invariant under infinite blueshifting of
scale? If there is new physics at some short distance scale,
whether it be the Planck scale or something longer, then the
gravitational redshift may lead to a communication
from short to long distance scales outside the horizon. That is,
{\it the redshift effect leads to a breakdown of the usual separation
of scales}.

Thus it seems perfectly possible that the quantum state of the outgoing
field modes near the horizon might {\it not} be the ground state.
The precise state of these modes could reflect details of physics
at much shorter distances. For instance, there may be amplitudes
for the excited states that could only be calculated from a
knowledge of the short-distance theory. If this is the case, then
the spectrum of black hole radiation may be quite different from
that deduced by Hawking.\footnote{This has nothing to do with the
fact that for interacting fields, the spectrum of black hole
radiation will reflect the dressing and decay of the interacting
particle states. Rather, we are referring to a
difference in the state of the high frequency modes, before the
interactions have had their effect.}
For example, if one of these modes
were to emerge at the cutoff in an excited state, then the
emission in that mode would be a combination of the spontaneous
Hawking radiation, the stimulated emission, and the original
excitation.\footnote{Stimulated emission by black
holes is analyzed in Ref. \cite{Waldstim}.}
Thus the flux of energy at infinity would be greater than
the Hawking flux.

\subsection{Constraints on the stress-energy tensor}
\label{subsec:constraints}

In this subsection we will analyze the implications for the
stress-energy tensor of a violation of the ground state
boundary condition near the horizon. The goal is to
determine what restrictions energy considerations may
place on the form of the quantum state of the outgoing
modes near the horizon. If the components of $\la T_{\mu\nu}\ra$
in the free-fall frame become too large, then neglect
of the back-reaction is unjustified. I see no reason in
principle why this may not happen in actuality. It may
be that, in fact, the problem of quantum fields
interacting with gravity in a black hole spacetime
defies treatment which neglects the back-reaction or
which treats it as a small perturbation that produces
only slow evaporation of the black hole mass. However,
if this is the case, then the (static) method of analysis used
in this paper is inapplicable.

Under what conditions can the back-reaction be treated
as a small perturbation? From the semi-classical
Einstein equation $G_{\mu\nu}=8\pi l_P^2 \la
T_{\mu\nu}\ra$, we infer that the back-reaction will
be small provided the stress tensor components
in the free-fall frame near the horizon are small
compared with $l_P^{-2}$ times the typical curvature
components there, i.e.,
\begin{equation}
\la T_{\mu\nu}\ra\ll 1/l_P^2 M^2.
\label{inequality}
\end{equation}
In the Unruh or Hartle-Hawking states, one has $\la
T_{\mu\nu}\ra=O(M^{-4})$ in the free-fall frame near the
horizon, hence in that state the back-reaction is very small indeed
as long as the hole is much larger than Planck size.
In fact, one must increase the stress tensor by a
factor of order $(M/M_P)^2$ before the back-reaction
becomes more than a small perturbation. This leaves
alot of leeway in the form of the state near the
horizon, and demonstrates that even within the
approximation that treats the back-reaction as a small
perturbation, there is no particular reason why the
ground state boundary condition at the horizon should
hold.

This leeway in the state at the horizon does not
necessarily mean that the black hole flux
would differ significantly from the Hawking flux
however. The
reason is that the energy carried by outgoing modes
near the horizon is vastly redshifted by the time they
make it out far from the hole. In order to make a
significant difference in the flux at infinity, an
excited outgoing mode near the horizon must have a very
high energy with respect to the free-fall frame.

To obtain a very crude estimate of the energy density
associated with such an excited mode, consider a wavepacket
that far from the hole is centered on a frequency $\o$
with a width $\Delta\o\sim\o$ and a spread in
retarded time $\Delta u\sim\o^{-1}$.
Suppose this mode is occupied in a one
particle state near the
horizon at some $r$. As discussed in section
\ref{subsec:existence},
its energy relative to the free-fall frame
will be roughly $(1-{2M\over r})^{-1}\; \o$, and the
proper volume of the thin spherical shell containing it will
be roughly $(1-{2M\over r})\; \o^{-1}M^2$ (since it has a thickness
$\Delta u\sim \o^{-1}$ at infinity).
Thus the energy density will be
roughly $(1-{2M\over r})^{-2}\; \o^2 M^{-2}$.\footnote{As discussed
in section \ref{subsec:existence}, the finite width of the
wavepacket leads to a differential redshift across the
packet, so this simple analysis is too crude to produce
reliable numerical coefficients.}
If this mode is followed all the way back to the
horizon, the energy density diverges, and the neglect
of the back-reaction is totally unjustified. If on the
other hand the mode is follwed only back to the value
of $r$ for which the $r$-frequency is equal to the cutoff
$\o_c$, then one has $(1-{2M\over r})^{-1}\sim \o_c/\o$,
and the energy density is $1/l_c^2M^2$.
Note that this result is
independent of $\o$, even though the extra power emitted
$\o/\Delta u\sim\o^2$ is not.

Now if the cutoff represents not just an arbitrary scale
beyond which we are pleading ignorance, but is rather a
physical scale at which the nature of propagation might
fundamentally change, then it might make sense to halt the
backward-in-time propagation when the $r$-frequency reaches
$\o_c$. Let us entertain this possibility.

Suppose then that the energy density near the horizon due to
the presence of an extra particle in the black hole
radiation is given by $1/l_c^2M^2$ as suggested by the
above computation. More extra particles would just multiply
this by the number of particles, irrespective of their
frequency.\footnote{The estimated energy density breaks down
however if the frequency is too low, because $\Delta u\sim
\o^{-1}$ will become so broad that the differential redshift
across the wavepacket totally invalidates the assignment of
a particular $r$-frequency to the packet near the horizon.}
(Note however that in order not to over-count
the degrees of freedom the independent modes should be spaced in
frequency by the spread adopted above, $\Delta\o\sim\o$.)
Similarly, for each particle {\it missing} from the Hawking
flux, one expects a {\it negative} contribution to the
energy density of the same magnitude.

Now if the back-reaction is a large effect, then our
analysis on the static black hole background is actually not
correct. We see no reasoning by which this scenario can be
ruled out, but we can say nothing more about it.
If on the other hand the back-reaction is small, then at least one of the
following must be true:
\begin{enumerate}
\item The outgoing modes have only small amplitudes to
be not in their ground state.
\item There is near-perfect cancellation between the
energy densities due to ``over-occupied" and
``under-occupied" modes.
\item The cutoff length $l_c$ is much longer than the
Planck length.
\end{enumerate}

It is not even entirely clear that (1) is consistent with a small
back-reaction, since it only implies a small {\it expectation
value} for the energy density, but still allows {\it
fluctuations} of order $1/l^2M^2$. It would seem to require
a full quantum theory of gravity to determine whether or not the
back-reaction could really be neglected in such circumstances.
While (2) cannot be ruled out, it seems somewhat implausible,
since there is no apparent reason for such cancellation to
occur. Also (3) does not seem very likely,
since there is currently no evidence of any
fundamental length scale other than the Planck length.
Nevertheless, let us just accept
these as the logical possibilities that they are. Is there
any further difficulty with such a scenario of deviation
from the Hawking spectrum maintaining small back-reaction?

If the {\it spectrum} of radiation is different, but the {\it luminosity}
is the same as the Hawking luminosity, then there must be cancellations
of positive and negative energy contributions, as mentioned in item
(2) above. Although this scenario does not seem likely, there seems
to be no way to rule it out. The possibility that the net luminosity
differs from the Hawking luminosity appears to be somewhat constrained
however by general properties of the stress energy tensor if the
back-reaction is to remain small.

As first shown in the 70's \cite{DavFullUn,ChristFull},
given some relatively ``theory-independent" constraints on
the behavior of the stress-energy tensor one can derive
a formula for the net radiation flux far from a (quasi)static black hole.
These constraints are:
\begin{itemize}
\item $\la T_{\mu\nu}\ra_{;}^{\nu}=0$;
\item $\la T_{\mu\nu}\ra$ is nonsingular on and outside
the horizon (in regular coordinates);
\item $\la T_{\mu\nu}\ra$ is static and spherically
symmetric (in four dimensions);
\item no radiation is incoming from infinity at late
times.
\end{itemize}
If all these properties hold then it can be shown\cite{ChristFull}
that the luminosity $L$ of the black hole is given in
two spacetime dimensions by
\begin{equation}
L={1\over2}\, M\, \int_{2M}^\infty dr\;
r^{-2} \la T_\alpha^\alpha\ra\qquad(D=2)
\label{lum2}
\end{equation}
and in four dimensions by
\begin{equation}
L=2\pi\, M\, \int_{2M}^\infty dr\; \la T_\alpha^\alpha\ra
+4\pi\int_{2M}^\infty dr\; (r-3M)\,
\la T_\theta^\theta\ra\qquad(D=4).
\label{L}
\end{equation}

Let us consider first the two-dimensional case. Then the
luminosity is determined entirely by the trace of the
stress-energy tensor. If we consider a conformally invariant
massless scalar field, the trace is determined in a
state-independent manner by the trace anomaly to be
$\la T_\alpha^\alpha\ra =R/24\pi$ where $R$ is the Ricci
scalar. Putting this in (\ref{lum2}) yields the Hawking
flux $L_H=1/768\pi M^2$.

Any deviation from the Hawking flux
for a conformally invariant field in two dimensions thus implies
that at least one of the properties of the stress tensor assumed
above must fail to hold. It seems that the most questionable assumption
is that of the value of the trace. But what would be the physical
basis for a deviation from the usual trace anomaly formula?

It was argued in \cite{Jac} that the presence of a high
frequency cutoff $\o_c$ is only likely to affect the
value of the trace by terms of order $O(R/\o_c^2)$.
This argument was based on the assumption that the
origin of the quantum violation of conformal invariance
can be located entirely in the regulated functional
measure in the manner of Fujikawa \cite{Fuji}.
If correct this implies that the corrections to the
trace (and to the Hawking flux) are very small indeed for
holes much larger than the cutoff length.
However, if there is fundamentally new physics at the
cutoff scale, then the violation of conformal
invariance will not be due simply to the non-invariance
of the regulated functional measure. This opens up the
possibility of a more significant deviation from the usual
trace anomaly. Nevertheless, the fact that the usual trace
anomaly is state-independent (assuming the state has the usual short
distance form down to some cutoff much smaller than the
radius of curvature of the spacetime) suggests strongly
that no significant deviation from the usual trace would occur.
Thus, at least in this two-dimensional model, it is hard to see
how the flux could differ from the Hawking flux and still have a small
back-reaction.

In the four dimensional case (\ref{L}) the situation is
perhaps different. For a conformally invariant field the trace is
still determined by the trace anomaly, and is given by
\begin{equation}
\la T_\alpha^\alpha\ra =\beta C^2/48=\beta M^2/r^6,
\label{trace}
\end{equation}
where $60\pi^2\beta=1,\; {7\over 4},\;
33/60\pi^2$ for fields of spin-0, 1/2 and 1
respectively, and $C^2$ is the square of the Weyl tensor.
Now however, the trace does not suffice to determine the luminosity.
One free function of $r$ remains undetermined.
The reason is that, unlike in two
dimensions where all metrics are conformally flat, the
Schwarzschild spacetime is {\it not} conformally flat,
so even a conformally coupled field scatters in a
non-trivial way. Both the spin of the field and the
detailed radial dependence of the metric affect the
radial dependence of $\la T_{\mu\nu}\ra$ and the net
flux at infinity.
Numerical computations \cite{Jensen} show, for example, that for
a massless, minimally coupled scalar field in the Unruh
vacuum in Schwarzschild spacetime,
the contribution of the second integral to (\ref{L})
is relatively small, and the Hawking luminosity is of order
$L_H\sim (4800\pi M^2)^{-1}\sim 10^{-4}M^{-2}$.

A deviation from the Hawking luminosity could be produced,
as in the two dimensional case, by a deviation from the
usual trace anomaly, however the same arguments as given
in that case make this seem unlikely. But in four dimensions
there is another possibility: Any change in
the tangential stress $\la T_\theta^\theta\ra$ will entail
a change in the luminosity $L$, without violating the above
assumptions on the behavior of $\la T_{\mu\nu}\ra$.
Can this be exploited to allow for a deviation from the Hawking
luminosity? While it is not clear why there should be any fundamental
difference between the two and four dimensional cases with regard
to the possibility of deviating from the Hawking radiation, let us
just take the result (\ref{L}) and see what can be done with it.

Note first that the $r$-dependence of $\la T_\theta^\theta\ra$
has alot to do with the scattering behavior of fields propagating in
the Schwarzschild geometry. Thus at most, we should think of the
possibility of freely modifying $\la T_\theta^\theta\ra$ at one
point, letting the behavior everywhere else be determined by
the scattering off the background geometry.
A change in the luminosity of order $\delta L$
could be produced, consistent with (\ref{L}),
in two qualitatively different
ways: (i) a change $\delta T_\theta^\theta\sim
O(\delta L/M^{-2})$ over a range $\delta r\sim M$,
or (ii) a very large change $\delta T_\theta^\theta$
over a very small range of $r$ near the horizon.
The second way seems inconsistent with the scattering
behavior of the field, since the effective potential that governs
the scattering is well behaved near the horizon. The first way
requires only a relative change
$\delta T_\theta^\theta/T_\theta^\theta$
of order unity to change the luminosity by order unity.
Thus there seems to be no obstacle to the physics at the cutoff
scale leading to a deviation from the Hawking luminosity, even if
the back-reaction is to remain small.

\section{Conclusion}
\label{sec:7}

What has been accomplished in this paper? We have
succeeded in formulating a derivation of the Hawking
effect (for massless free fields) that avoids reference
to field modes above some
cutoff frequency in the frame of the free-fall
observers that are asymptotically at rest. To stay
below the cutoff it is necessary to impose a
boundary condition on the field  near the horizon {\it for all
times}. The boundary condition states roughly that the
outgoing high frequency field modes are in their ``ground
state" as viewed by free-fall observers.
This boundary condition is not derivable from
the initial state within the cutoff theory.

The precision of our derivation is controlled by the
ratio of the cutoff length to the Schwarzschild radius
of the black hole, and is limited by $\sqrt{l_c/M}$.
For a black hole large compared with
the cutoff length, the largest source of imprecision
is the unavoidable spread of the wavepackets employed,
and the associated large differential in the redshift suffered
across the packet when it is near the horizon.

The boundary condition we impose may
or may not be physically the correct one.
If it fails to hold, then there will be a deviation from
the Hawking spectrum. It seems this could occur either with or without
a large back-reaction. If the back-reaction is to remain small,
then either the deviation must be small, or there must be cancellation
between positive and negative energy contributions, or there must
be a physical cutoff much longer than the Planck length.
In four dimensions, the generally expected behavior of the
stress-tensor cannot be used to definitively
rule out any of these scenarios.

Even a very small deviation from the
thermal nature of the Hawking radiation would seem to entail a
breakdown in the generalized second law of
thermodynamics \cite{Beck,UnWald,ThorneZurek}.
Thus one has reason to suspect that the
physics at the cutoff scale somehow conspires to
produce precisely the ``thermal" state. However, that is not
to say that the ordinary effects of quantum field
propagation in the black hole background should not
leave their mark on the radiation. The scattering of
wavepackets by
the geometry is one well-known aspect of this mark,
but it is conceivable that the redshifting of the
physics at the cutoff is another one. If that is the
case, then the thermodynamic behavior of physics in a
black hole spacetime may turn out to be much more
subtle than was previously thought.

Given a candidate theory with a short distance cutoff,
it will certainly be interesting to study its behavior
in a black hole spacetime, in which the redshift
effect acts as a microscope to reveal consequences of
short-distance physics at larger scales.

\section*{Acknowledgments}

I am grateful to John Dell, Kay Pirk and Jonathan Simon for several
helpful discussions. This research was supported by
the National Science Foundation under Grant Nos. PHY91-12240
and PHY89-04035.

\section*{Appendix A: Black hole line element}

The static, spherically symmetric black hole line element
in Schwarzschild, tortoise, double-null, and ingoing
Eddington-Finkelstein coordinates takes the following forms:
\begin{eqnarray} ds^2&=
&(1-{\textstyle{2M\over r}}) dt^2
-(1-{\textstyle{2M\over r}})^{-1} dr^2
-r^2d\Omega^2\\
&=&(1-{\textstyle{2M\over r}}) (dt^2-d{r^*}^2)-r^2d\Omega^2\\
&=&(1-{\textstyle{2M\over r}}) dudv- r^2d\Omega^2\\
&=& (1-{\textstyle{2M\over r}}) dv^2-2 dvdr-r^2d\Omega^2
\end{eqnarray}
with
\begin{equation}
r^*=r+2M\ln({\textstyle{r\over2M}}-1)
\label{tortoise}
\end{equation}
\begin{equation}u=t-r^*,\qquad\qquad v=t+r^*.
\label{uv}
\end{equation}
The coordinates $u$ and $v$ are called the {\it retarded} and
{\it advanced} time coordinates respectively.

If $x^{\mu}(\lambda)$ is an affinely parametrized
geodesic, then it is a stationary point of the integral
$\int g_{\mu\nu}\dot{x}^{\mu}\dot{x}^{\nu} d\lambda$, where
the dot $\cdot=d/d\lambda$.
To see that $r$ is an affine parameter along ingoing radial
null geodesics, it is convenient to use the
Eddington-Finkelstein coordinates, so that
$\dot{v}=\dot{\theta}=\dot{\phi}=0$. Upon varying
$v(\lambda)$, one immediately finds $\ddot{r}=0$,
so $r=a\lambda+b$ for some constants $a$
and $b$.

\section*{Appendix B: Klein-Gordon inner product}

The Klein-Gordon inner product $\la f,g\ra$ between two
initial data sets $f$ and $g$ on a Cauchy
surface $\Sigma$ is defined by
\begin{equation}
\la f,g\ra=\int j^\mu \, d\Sigma_{\mu}
\label{B7a}
\end{equation}
\begin{equation}
j^\mu= {\textstyle {i\over2}}\sqrt{-g}g^{\mu\nu}(f^*\partial_\nu
g-g\partial_\nu f^*).
\end{equation}
The surface element $d\Sigma_\mu$ is given by
\begin{equation}
d\Sigma_\mu={1\over6}\epsilon_{\mu ijk}\, d\sigma^i
d\sigma^j d\sigma^k,
\end{equation}
where $\sigma^i$ ($i=1,2,3$) are coordinates on the surface
$\Sigma$. For solutions of the KG equation of compact
support, (\ref{B7a}) is independent of the Cauchy surface on which
the integral is evaluated, since the current vector density
$j^\mu$ is divergence free, $\partial_\mu j^\mu=0$.
We shall have occasion to evaluate the KG inner product on a
surface that is null, which can be thought of as a limiting
case of Cauchy surfaces.

\section*{Appendix C: Quantum field theory}

The field operator $\Phi$ for a real, free scalar field
is a Hermitian operator that satisfies
the wave equation $\nabla^2\Phi=0$. We define an
annihilation operator corresponding to an initial data set
$f$ on a surface $\Sigma$ by
\begin{equation}
a(f)=\la f,\Phi\ra_\Sigma.
\label{C2}
\end{equation}
If the data $f$ is extended to a solution of the wave
equation then we can evaluate the KG product in (\ref{C2}) on
whichever surface we wish. The hermitian adjoint of $a(f)$
is called the creation operator for $f$ and it is given by
\begin{equation}
a^{\dagger}(f)=-\la f^*,\Phi\ra_\Sigma.
\end{equation}
The commutation relations between these operators follow
from the canonical commutation relations satisfied by the
field operator. The latter are equivalent to
\begin{equation}
 [a(f),a^{\dagger}(g)]=\la f, g\ra,
\label{C3}
\end{equation}
provided this holds for all choices of $f$ and $g$.
Now it is clear that only if $f$ has positive, unit KG norm
are the appelations ``annihilation" and ``creation"
appropriate for these operators.
{}From (\ref{C3}) and the definition of the KG inner product
it follows identically that we also have the commutation
relations
\begin{equation}
[a(f),a(g)]=-\la f, g^*\ra,\qquad
[a^{\dagger}(f),a^{\dagger}(g)]=-\la f^*, g\ra.
\end{equation}

A Hilbert space of ``one-particle states" can be defined by
choosing a decomposition of the space $S$ of complex
initial data sets (or solutions to the wave equation) into a
direct sum of the form $S=S_p\oplus S_p{}^*$, where all the
data sets in $S_p$ have positive KG norm and the space $S_p$
is orthogonal to its conjugate $S_p{}^*$. Then all of the
annihilation operators for elements of $S_p$ commute with
each other, as do the creation operators. A ``vacuum" state
$|\Psi\ra$ corresponding to $S_p$ is defined by the condition
$a(f)|\Psi\ra=0$ for all $f$ in $S_p$, and a Fock space of
multiparticle states is built up by repeated application
of the creation operators to $|\Psi\ra$.

(Instead of thinking of the Hilbert space as the Fock space
corresponding to some decomposition $S_p\oplus S_p{}^*$ as
above, it is perhaps conceptually preferable to take the
point of view of the algebraic approach to quantum field
theory \cite{KayWald}, according to which a ``state" is
simply a positive
linear functional $\rho$ on the $\star$-algebra of field
operators. Thus for example, to express the idea that a
given field mode $f$ is in its ground state, one says that
the state $\rho$ satisfies $\rho({\cal O} a(f))=0$ for all
operators ${\cal O}$. This language is preferable if, as is
often the case for quantum fields in curved space, one
wishes to simultaneously consider a state as an element of
two completely differently constructed (for example ``in" and
``out") Fock spaces. In the
algebraic approach, no mysterious ``identification" of the
two Fock spaces is required. Another advantage is that
whereas the statement that the field operator is ``hermitian" is
meaningless until the Hilbert space on which it acts has
been specified, the statement that $\Phi$ goes into itself
under the abstract $\star$ operation is always well defined.
The algebraic approach is clearly preferable in
contexts (e.g. \cite{KayWald,WaldYurt}) in which one wishes to
obtain results valid for a class of quantum states that is
as wide as possible.)

\section*{Appendix D: Negative frequency part \\
of the transmitted wavepacket}

Consider the transmitted part $t_{\obar\ubar}$ of a
wavepacket $p_{\obar\ubar}$ propagating in the
Schwarzschild black hole spacetime, narrowly peaked in
$u$-frequency about $\obar$ (at large $r$)
and about some late retarded
time $\ubar$. For the original Hawking argument one needs to
determine the KG norm of the negative frequency part of
$t_{\obar\ubar}$ at $\pastI$ in terms of the norm of
$t_{\obar\ubar}$ itself. For our argument in this paper,
it is the negative $r$-frequency part on a constant $v$ surface
that is of interest. It was Hawking's original argument that
these two are related, using the geometrical optics
approximation to propagate the very high frequency modes in
question back out to $\pastI$.

Consider a collection of null surfaces, wavefronts for such a mode.
In Fig. 2 one surface is shown that is outgoing at
retarded time $u$ and ingoing at advanced time $v(u)$. The
key fact is that for late retarded times,
the value of the affine parameter $r(u,v_c)$ where this
wavefront intersects the surface $v=v_c$ is linearly
related to the advanced time $v(u)$.\footnote{Hawking
argued that this is because as one goes from $(u,v_c)$ back
along the wavefromt and out to $\pastI$, the ``vector" that
connects the wavefront to the horizon (and earlier to the null
ray that becomes the generator of the horizon) is parallel
transported into itself. This is not actually correct, since
the connecting vector satisfies not the parallel transport
equation but the geodesic deviation equation. Nevertheless,
one still obtains a finite linear scaling of the connecting vector,
which is all that is required for the argument \cite{WaldGR}.}
Therefore the negative
$r$-frequency part of $t_{\obar\ubar}$ at $v=v_c$ propagates
back to the negative $v$-frequency part at $\pastI$. Thus
the corresponding KG norms are identical, so in both cases
we can carry out the calculation at $v=v_c$.

Now there is an observation \cite{Notes,Wald75} that makes the
extraction of the negative frequency part simple: let
$U$ be defined by $\kappa u=-\ln(-\kappa U)$, and consider
the functions $q$ and $\widetilde q$, defined by
\begin{equation}
q(U)=\left\{
\begin{array}{l}
e^{-i\omega u} \quad {\rm for}\;  U<0\\
0\quad\quad\; \; {\rm for}\;  U>0
\end{array}\right.
\end{equation}
and
\begin{equation}\widetilde{q}(U)=q(-U).
\end{equation}
That is, $\widetilde q$ is just the function $q$ reflected over
the line $U=0$ ($u=\infty$). Then one can easily show that
the functions
\begin{eqnarray}
q^{\scriptscriptstyle(+)}&=&c_{+}
(q+e^{-\pi\o/\kappa}\widetilde{q})
\qquad {\rm and}\label{q+}\\
q^{\scriptscriptstyle(-)}&=&c_{-}
(e^{-\pi\o/\kappa}q+\widetilde{q})
\end{eqnarray}
are pure positive and negative $U$-frequency packets
respectively. One can solve for the normalization factors
$c_{+}$ and $c_{-}$ by setting
$q=q^{\scriptscriptstyle(+)}+q^{\scriptscriptstyle(-)}$.
This yields
\begin{eqnarray}
c_{-}&=&-e^{-\pi\o/\kappa}\; c_{+},\\
c_{+}&=&(1-e^{-2\pi\o/\kappa})^{-1}.
\end{eqnarray}
Finally, the KG norm of $q^{\scriptscriptstyle(-)}$ is
calculated from (54) and (55,56) using
$\la\widetilde{q},\widetilde{q}\ra=-\la q,q\ra$ and $\la
q,\widetilde{q}\ra=0$, yielding
\begin{equation}
\la q^{\scriptscriptstyle(-)},q^{\scriptscriptstyle(-)}\ra
=-\la q,q\ra (e^{2\pi\o/\kappa}-1)^{-1}.
\end{equation}

The preceeding calculation is directly applicable to the
wavepacket $t_{\obar\ubar}$, squeezed near the horizon on
the surface $v=v_c$.
The relation between $r$ and $u$ along $v=v_c$ is given by
(42,43), $u=v_c-2r^*=v_c-2r-4M\ln({r\over2M}-1)$. For our
wavepacket near the horizon, the spread in $r$ is very small
compared with $2M$, so the wavepacket only has support where
one has $\kappa u\simeq-\ln({r\over 2M}-1)+{\rm const.}$.
Thus $U$ and $r$ are linearly related
(via $-\kappa U={r\over 2M}-1$), so the negative $r$-frequency
part $q^{\scriptscriptstyle(-,r)}$ is equal to the negative
$U$-frequency part $q^{\scriptscriptstyle(-)}$ (54) with
$\o=\obar$, provided the packet is sufficiently peaked in
frequency about $\obar$ ($\Delta\o\ll\kappa$) so that the
expressions (53,54) for the positive and negative frequency
parts still hold.
This is the result used in the text.


\section*{FIGURE CAPTIONS}

\noindent Figure 1. Conformal diagram depicting wavefronts of the
wavepacket $P=R+T$ propagating in the spacetime of a
spherically symmetric collapsing body.
\vspace{1cm}

\noindent Figure 2. Conformal diagram depicting the propagation of a wavefront.
The point $(v_c, r(u,v_c))$ is connected by a radial null geodesic
to a point on $\pastI$ at advanced time $v(u)$. The affine parameter
$r$ along the line $v=v_c$ is linearly related to $v(u)$ for late
retarded times $u$.

\end{document}